# STATISTICALLY OPTIMAL MODELING OF FLAT ECLIPSES AND EXOPLANET TRANSITIONS. THE "WALL-SUPPORTED POLYNOMIAL" (WSP) ALGORITMS

*Andrych K.D.*[1,2], *Andronov I.L.*[2], *Chinarova L.L.* [3,2]

[1] Department of Theoretical Physics and Astronomy, Odessa National University
[2] Department of Mathematics, Physics and Astronomy, Odessa National Maritime University
[3] Astronomical Observatory, Odessa National University
*katyaandrich@gmail.com, tt_ari@ukr.net, llchinarova@gmail.com*

**ABSTRACT.** The methods for determination of the characteristics of the extrema are discussed with an application to irregularly spaced data, which are characteristic for photometrical observations of variable stars. We introduce new special functions, which were named as the "Wall-Supported Polynomial" (WSP) of different orders. It is a parabola (WSP), constant line (WSL) or an "asymptotic" parabola (WSAP) with "walls" corresponding to more inclined descending and ascending branches of the light curve. As the interval is split generally into 3 parts, the approximations may be classified as a "non-polynomial splines".

These approximations extend a parabolic/linear fit by adding the "walls" with a shape, which asymptotically corresponds to the brightness variations near phases of the inner contact. The fits are compared to that proposed by Andronov (2010, 2012) and Mikulasek (2015) and modified for the case of data near the bottom of eclipses instead of wider intervals of the light curve. The WSL method is preferred for total eclipses showing a brightness standstill. The WSP and WSAP may be generally recommended in a case of transit eclipses, especially by exoplanets. Other two methods, as well as the symmetrical polynomials of statistically optimal order, may be recommended in a general case of non-total eclipses.

The method was illustrated by application to observations of a newly discovered eclipsing binary GSC 3692-00624 = 2MASS J01560160+5744488, for which the WSL method provides 12 times better accuracy.

**Keywords:** *variable stars; eclipsing binary; minima timings; O-C analysis; TYC 3692-624-1 = Gaia DR1 505352827074254080 = GSC 03692-00624 .*

## 1. Introduction

The O-C analysis is the most popular method of studies of period variations (cf. Tsesevich 1970, 1971, Kreiner et al. 2001, Andronov et al., 2017). Many astronomers observe stars only during relatively short intervals near extrema (minima of eclipsing binaries or maxima of the pulsating variables) instead of the complete light curves.

Currently in the AAVSO (2017) and BAA (2011) user guides, the Times of Minima (or Maxima, the letter "M" is the same) are abbreviated to "ToM". This needs adequate methods of modeling, which will provide best quality of approximation.

In the pre-computer era, the most popular was method of chords by Pogson, where the approximation of points on the graph was made manually. More advanced method was proposed by Herzsprung, where the mean curve was estimated by binning the data in to time (or phase) intervals and then linearly interpolated. Such a curve was shifted and scaled to the particular data to obtain individual extrema timings.

These historical methods were discussed in numerous monographs and textbooks (e.g. Tsesevich 1970, 1971).

In the computer era, one may expect to make "physical" modeling using the code based on the method by Wilson & Devinney (1971) and its improvements (e.g. Zoła, Kolonko & Szczech,1997, Zoła et al., 2010, Prša & Zwitter, 2005). Or to use a simplified physical model (e.g. Andronov & Tkachenko 2013).

However, the number of the parameters in these models is still too large. So one may get approximations of nearly the same quality for a relatively large region in the parameter space. Thus "phenomenological" approximations are still valid for a dominating majority of variable stars.

Andronov (1987) elaborated software based on periodic cubic splines, which allows determination of the local or global approximations and, particularly, of the characteristics of extrema. More functions were reviewed by Andronov (2005).

Specially for (generally) asymmetric maxima of pulsating variables, Marsakova & Andronov (1996) proposed an algorithm of "Asymptotic parabolae" (AP), which was used for a compilation of the catalogue of characteristics of extrema of a group of the Mira-type stars (Andronov & Marsakova, 2006).

However, Chinarova & Andronov (2000) used an algebraic polynomial approximation of a degree corresponding to a minimum error estimate of timing.

Previous methods were typically based on the assumption of smooth functions, whereas eclipses definitely show begin and end not only in EA-type (Algol) systems (Samus et al. 2017), but also in EB-type (β Lyr) and even EW-type (W UMa) systems (Andronov 2012ab, Tkachenko, Andronov & Chinarova, 2016).

Thus it is natural either to apply non-polynomial splines to a complete phase light curve, or to make local approximations near the extrema.

In this paper, we compare approximations using "old" and "new" sets of functions for statistically optimal modeling of symmetrical minima of eclipsing binary stars.

## 2. The Observations

The methods are illustrated by application to one "flat" minimum of the eclipsing binary GSC 3692-00624 = 2MASS J01560160+5744488 (Devlen, 2015). From the complete data set, we extracted a "full" (HJD 2455506.28847 – .43138, n=43) and "part" (HJD 2455506.32250 – .40077, n=24) intervals.

As previous algorithms were effective for "smooth" minima, here we concentrate on the "flat" minima characteristic for transits of stars and exoplanets.

## 3. The Methods
### 3.1. Asymmetrical polynomials

The basic method of the approximation is based on algebraic polynomials. They are included in popular electronic tables in the Microsoft, Open/Libre, Kingsoft offices, GNUmeric and others. However, these approximations are only shown on graphs, without a possibility of getting precise coefficients and error estimates of them.

Algebraic polynomials were implemented for the minima determination in the software VSCalc (Breus, 2006) and PERANSO (Vanmunster, 2015) with an user freedom to choose the degree of the polynomial. Generally, these polynomials are asymmetrical for $m > 3$.

Typically, the duration of observations is much less than the time from some "adopted" starting point (e.g. ad in the Julian Date (JD)). Thus it is suitable to rescale the time argument $t$ to the symmetrical interval $[-1,+1]$:

$$u = 2 \cdot (t - t_{n_1})/(t_{n_2} - t_{n_1}) - 1 \qquad (1)$$

Here $t_{n_1}$ and $t_{n_2}$ are times of the beginning and end of the interval of observations. Thus

$$x_P(t) = \sum_{\alpha=1}^{m} C_\alpha \cdot u^{\alpha-1} \qquad (2)$$

These functions are generally asymmetrical in respect to the moment of extremum, thus may be used for studies of pulsating variables as well.

In our program MAVKA (Multi-Analysis of Variables by Kateryna Andrych), we have used various approximations, not only the ordinary algebraic polynomials.

The preliminary version was introduced by Andrych et al. (2015).

### 3.2. Symmetric polynomials

Another set of functions on time $t$ are symmetric algebraic polynomials

$$x_{SP}(t) = \sum_{\alpha=1}^{m} C_\alpha \cdot v^{2(\alpha-1)} \qquad (3)$$

for different $m$. Here $v = u - u_e$ is the shift from the argument of symmetry $u_e = C_{m+1}$.

Because of this specific form, the symmetric polynomial defined by $(m+1)$ parameter is of the degree $2(m-1)$ instead of $m$ for the generally asymmetrical algebraic polynomial. So the parabola (degree 2) is defined by 3 parameters in both cases. But next steps in the degree of 2 require 1 or 2 parameters, respectively.

Such an approximation coincides exactly with an algebraic polynomial of order 0 (constant, $m=1$) and 2 (parabola, $m=2$). The parameters $C_1..C_m$ are determined using the LS (Least Squares) method with further improvement of $C_{m+1}$ using differential corrections (cf. Andronov 1994, 2003). Numerical experiments show that sometimes there may be "inverse" variations, typically at the borders and at a mid-eclipse, if observationally it is flat enough.

The symmetrical polynomial fits are shown in Fig. 1 for different number of parameters. The $\pm 1\sigma$ "error corridors" for $x_{SP}(t)$ are shown as well, but are typically comparable to the thickness of line. As expected, the increase of the number of parameters leads to smaller deviations of the data from the fit, but the minimum becomes split. Thus the point of symmetry $u_e$, which should correspond to the minimum, formally corresponds to the maximum brightness surrounded symmetrically by deeper local minima.

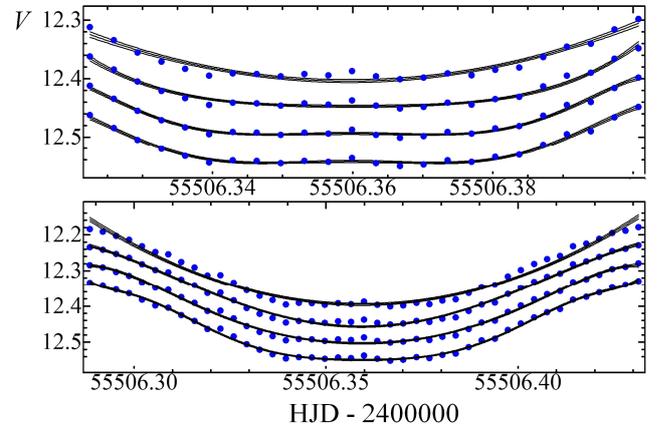

Figure 1: "Symmetric polynomial" approximations for "part" (up) and "full" (down) intervals of the same eclipse. The approximations are shown for the number of parameters $m = 3,4,5,6$ (from up to down) with an artificial shift of $0.05^m$.

### 3.3. Non-integer degrees as asymptotic fits

Andronov, Tkachenko & Chinarova (2017) tested many functions and modifications and ranged them according to the quality of approximation. There are two families of functions describing the shape (synonyms: "pattern", "form") of the eclipse. The first suggests a limited width of the eclipse (Andronov, 2010, 2012a)

$$G_A(z) = \begin{cases} 1 - (1 - |z|^\beta)^{1.5}, & \text{if } |z| \le 1, \\ 0, & \text{if } |z| > 1, \end{cases} \qquad (4)$$

where $z = (u - u_e)/\Delta = v/\Delta$ is dimensionless time, $u_e$ is again an argument of extremum, $\Delta$ is the eclipse half-width. This approximation is called NAV ("New Algol Variable").

The second family corresponds to a formally unlimited width (Mikulášek et al., 2011, Brat et al., 2011):

$$G_M(z) = (1 - \exp(-\vartheta))^\gamma, \quad (5)$$

where $\vartheta(z) = z^2/2$ corresponds to a Gaussian function for $\gamma = 1$. Despite existence of advanced functions, this simplest function was used recently for studies of GAIA observations of eclipsing variables (Mowlavi et al., 2017).

An improved function (Mikulášek, 2015) may be decomposed into Mac-Laurin series (Andronov et al., 2016):

$$\vartheta(z) = \cosh(z) - 1 = \frac{z^2}{2} \cdot (1 + \frac{z^2}{12} + \frac{z^4}{360} + ...) \quad (6)$$

A restricted form of these functions is below:

$$G_A(z) = \frac{3}{2}|z|^\beta - \frac{3}{8}|z|^{2\beta} + ... \quad (7)$$

$$G_M(z) = \vartheta^\gamma \cdot (1 - \frac{\gamma\vartheta}{2} + \frac{(3\gamma^2 + \gamma)}{24}\vartheta^2 + ...) \quad (8)$$

At the bottom of the eclipse, far of its borders, the most important are the first terms. Thus the restricted approximations are

$$x_A(z) = C_1 + C_2|v|^\beta + C_3|v|^{2\cdot\beta} \quad (9)$$

$$x_M(z) = C_1 + C_2|v|^\beta + C_3|v|^{2+\beta} \quad (10)$$

So the difference is only in the third term. For the compatibility, we adopt $\beta = 2\gamma$. Hereafter we call these approximations as NAVs and Ms, respectively. Besides the three coefficients seen in the formulae, there are two more coefficients ($u_e = C_{m+1}$ and $\beta = C_{m+2}$), which are determined by differential corrections after searching for a minimum of the test function

$$\Phi = \sum_{k=n_1}^{n_2} w_k \cdot (x_k - x_C(u_k))^2 \quad (11)$$

at a grid. The total number of parameters here is $m_p = m + 2$. As we do not extrapolate, the minimum is expected to be inside of the selected interval of observations $t_e \in [t_{n_1}, t_{n_2}]$, $u_e \in [-1, +1]$.

The parameter $\beta \in [1,5]$. Formally, the upper limit may be set to infinity, but the listed value is enough for minimization on a grid, and may be corrected to (e.g.) larger values using differential corrections.

Similarly, for the family of functions

$$G_{SP}(z) = 1 - (1 - z^2)^\gamma \quad (12)$$
$$= \gamma z^2 + \frac{\gamma(\gamma-1)}{2}z^4 + \frac{\gamma(\gamma-1)(\gamma-2)}{6}z^6...$$

(Andronov, 2012b) one may get a subset of "symmetric polynomial" (SP) approximation with integer powers:

$$x_{SP}(z) = C_1 + C_2 u^2 + C_3 u^4 + C_4 u^6 + ... \quad (13)$$

The approximation (12) was recently applied by Juryšek et al. (2017) for studies of a large sample of eclipsing binaries with changing inclination in the LMC.

The unbiased estimate of the root mean squared (r.m.s.) deviation of the data from the fit

$$\sigma_0 = \sqrt{\frac{\Phi}{n - m_p}} \quad (14)$$

As will be shown below, these analytical functions produce apparent waves, like the Gibbs phenomenon in the trigonometric polynomial fits of the complete phase light curves (e.g. Andronov, Tkachenko & Chinarova, 2016). However, during the transits or total eclipses, it is natural to split the data at the intervals of inner contacts.

### 3.4. "Wall-Supported" Functions

This series of functions we call "wall-supported" (WS), assuming some symmetrical basic function $V(v)$ for the integral of total minimum, and another symmetrical function $W(v)$ outside.

We tested different approximations. The corresponding plots are shown in Figures 2 and 3 for "part" and "full" data sets, respectively. The smoothing functions are shown in Fig. 4 and 5.

Taking into account a limb darkening, we initially suggested a "Wall-Supported Parabola"

$$x_{WSP}(u) = C_1 + C_2 \cdot v^2 + C_3 \cdot W_+(|v| - \delta) \quad (15)$$

Here the "Wall" function of $\varsigma = |v| - \delta$

$$W_+(\varsigma) = \begin{cases} 0, & \text{if } \varsigma \leq 0 \\ W(\varsigma), & \text{if } \varsigma > 0 \end{cases} \quad (16)$$

Here $u_s, u_f -$ are the arguments of "start" and "finish", respectively. Obviously, the argument of extremum is $u_e = (u_s + u_f)/2$, $u_s = u_e - \delta$, $u_f = u_e + \delta$, and the half duration of the middle part $\delta = (u_f - u_s)/2$.

To determine "non-linear" parameters in each approximation, we computed of the test function $\Phi$ on a grid.

The lines of equal level of $\Phi$ at the two-parameter diagrams for the NAV method applied to a simplified physical model were presented by Tkachenko (2016).

To study test functions for our models, we preferred to change the type of presentation, taking into account a locally paraboloidal shape of the test function. Thus, in our program, we made a following routine for the 2D-plots. The test function was computed over a grid (201×201). Then we computed a dimensionless value

$$p = \sqrt{(\Phi - \Phi_{\min})/(\Phi_{\max} - \Phi_{\min})}, \quad (17)$$

which we expect to represent a set of equal levels better than (typically) without the square root. For an exact paraboloid, the constant step in $p$ leads to concentric ellipses with a constant step in each fixed direction from the center. The interval [0,1] for $p$ is split into $N_p = 10$ subintervals, inside each of them the color linearly changes from blue to yellow.

This makes an abrupt change of color between the subintervals. Such a kind of presentation allows to get advantages of the "gradient" and "lines" styles.

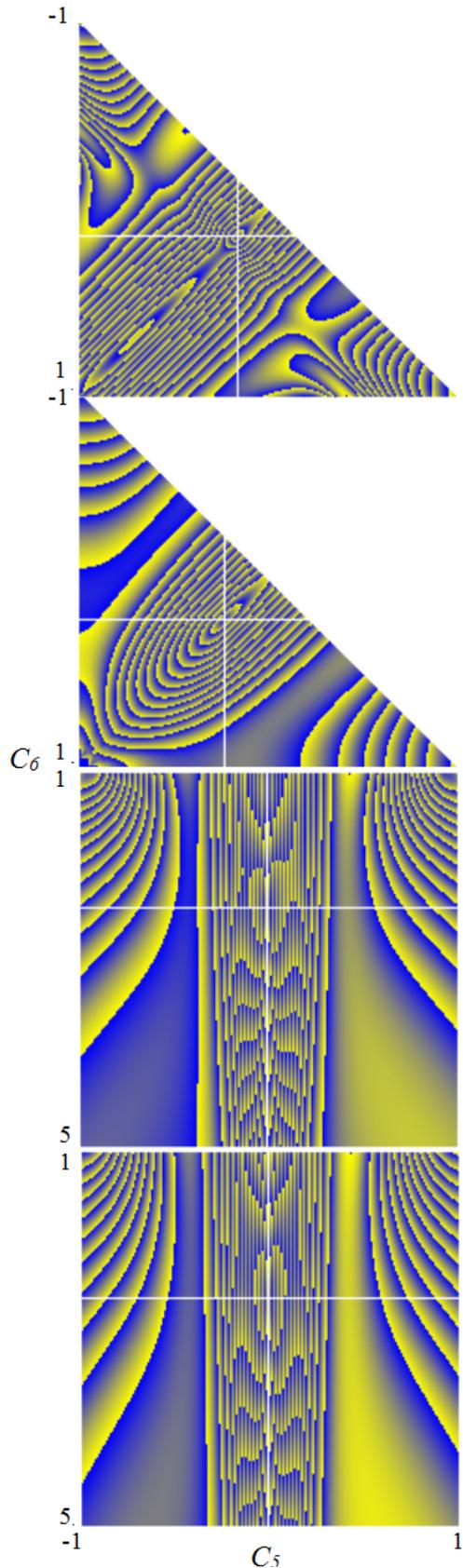
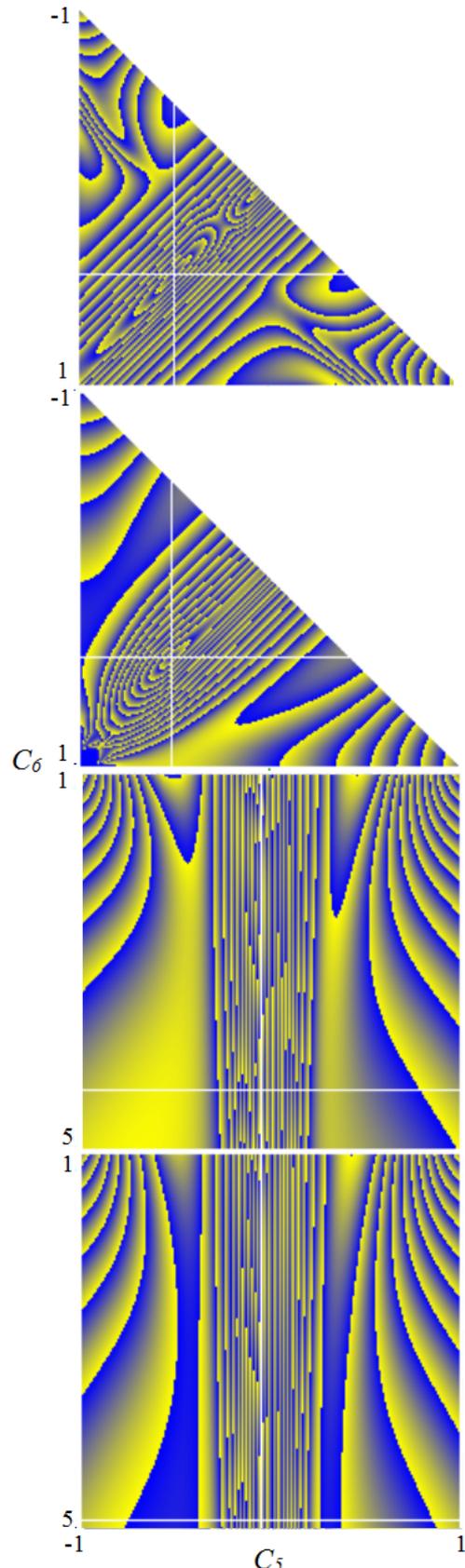

Figure 2: The dependence of the test function on the parameters $C_5$ ("scaled" time from $-1$ to $1$) and $C_6$, for "full" data, depending on the model (up to down): WSP, WSL ($C_5=u_s$, $C_6=u_f$), "NAVs" (Eq. (9)), "Ms" (Eq. (10)) ($C_5=u_e$, $C_6=\beta$). The white cross shows the minimum.

Figure 3: The dependence of the test function on the parameters $C_5$ ("scaled" time from $-1$ to $1$) and $C_6$, for "part" data, depending on the model (up to down): WSP, WSL ($C_5=u_s$, $C_6=u_f$), "NAVs" (Eq. (9)), "Ms" (Eq. (10)) ($C_5=u_e$, $C_6=\beta$).

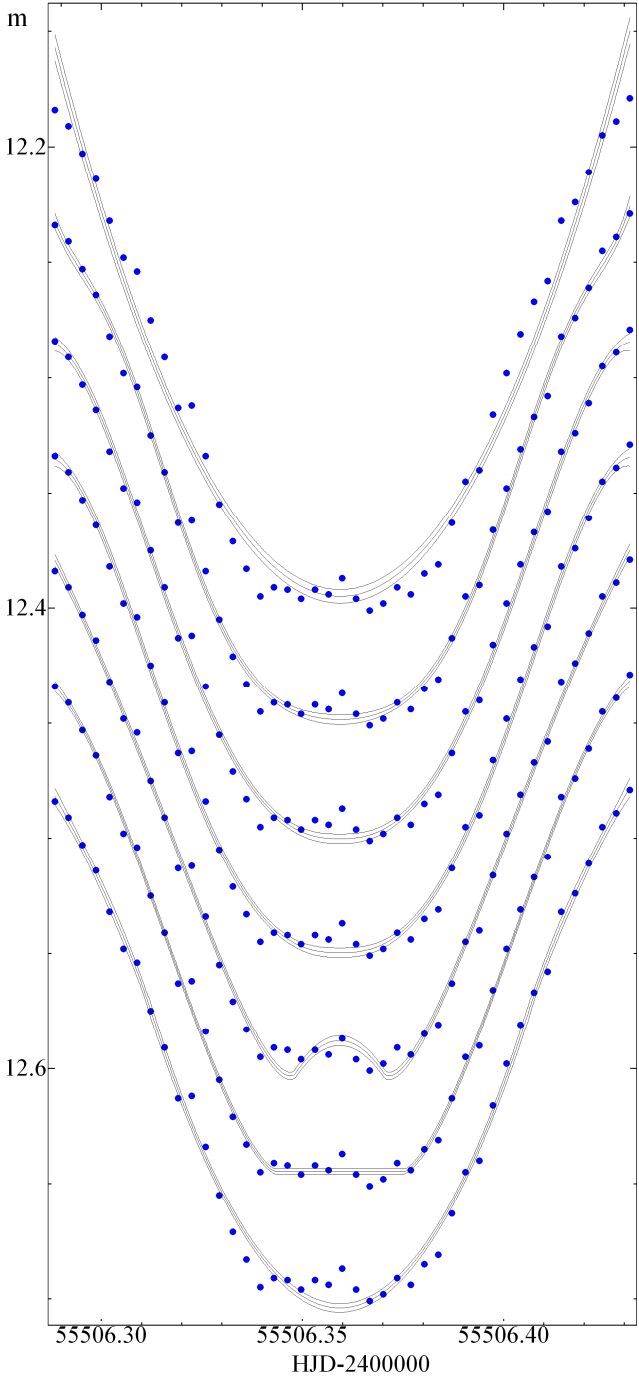
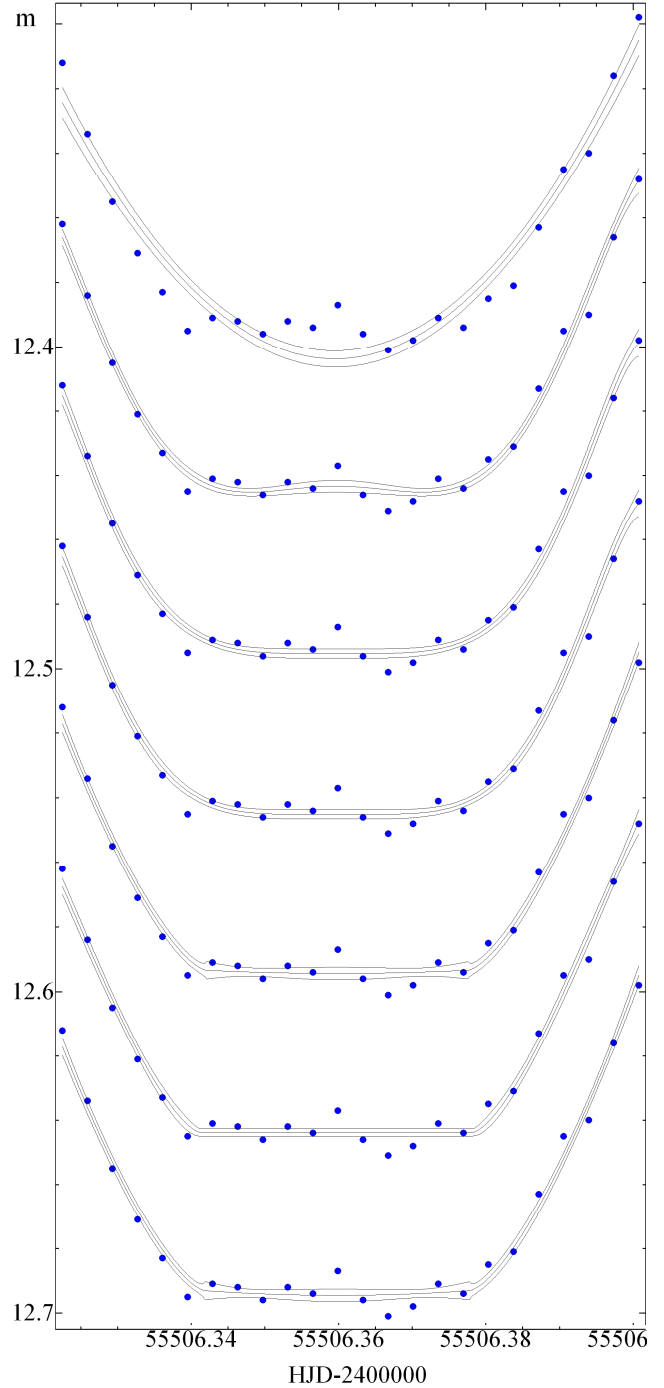

Figure 4: Approximations of the "full" data using the methods (up to down, as the graphs):

| $x_C$ | $t_e$ | ± | $m_e$ | ± | $\sigma_0$ | $m_p$ |
|---|---|---|---|---|---|---|
| P | 369 | 494 | 9484 | 296 | 1294 | 3 |
| SP | 925 | 240 | 9843 | 211 | 561 | 5 |
| NAVs | 259 | 268 | 10002 | 208 | 599 | 5 |
| Ms | 259 | 263 | 9959 | 206 | 588 | 5 |
| WSP | 234 | 220 | 8790 | 217 | 523 | 5 |
| WSL | 248 | **188** | 9450 | **134** | **426** | 5 |
| WSAP | 264 | 294 | 10393 | 187 | 663 | 5 |

With the bold font, we mark the minimal error estimates among the different approximations. Time $t_e$ is as $10^6$(HJD-2455506.359), brightness $m_e=10^5$ (V-12.3), the error estimates are scaled similarly.

Figure 5: Approximations of the "part" data using the methods (up to down, as the graphs):

| $x_C$ | $t_e$ | ± | $m_e$ | ± | $\sigma_0$ | $m_p$ |
|---|---|---|---|---|---|---|
| P | 504 | 637 | 10369 | 256 | 836 | 3 |
| SP | 1132 | 272 | 9326 | 183 | 416 | 5 |
| NAVs | 744 | 283 | 9517 | 145 | 437 | 5 |
| Ms | 734 | 281 | 9501 | 142 | 432 | 5 |
| WSP | 800 | 267 | 9431 | 193 | 425 | 5 |
| WSL | 684 | **272** | 9380 | **124** | **415** | 5 |
| WSAP | 800 | 266 | 9447 | 193 | 425 | 5 |

With the bold font, we mark the minimal error estimates among the different approximations. The scaling is as in Fig. 4.

As the "wall function", we initially adopted $W_1(\varsigma) = \varsigma^{1.5}$ as theoretically expected for the asymptotic behavior of the light curve near inner and outer contacts, except for the rare case of equal radii, when the power index is 1 for the inner contact (cf. Andronov, 2012a, Andronov & Tkachenko, 2013). More simple, than WSP, appeared a "Wall-Supported Line" (WSL) with a constant

$$x_{WSL}(u) = C_1 + C_2 \cdot (\varsigma^{1.5})_+ + C_3 \cdot (\varsigma^{3.5})_+ \quad (18)$$

Here again the index "+" means that the function is non-zero only for positive argument.

Finally, we introduce an "Asymptotic Parabola" (AP) (Marsakova & Andronov, 1996) base (Eq. 19):

$$x_{WSAP}(u) = \begin{cases} C_1 + C_2 \cdot v^2, & \text{if } v \leq \delta, \\ C_1 + C_2 \cdot (2|v| - \delta)\delta + C_3 \cdot \varsigma^{1.5}, & \text{if } v > \delta \end{cases}$$

Obviously, the dependence is on time, other variables $(u, v, \varsigma)$ are related and introduced for suitability.

### 4. Discussion

Previous methods were typically based on the assumption of smooth functions, whereas eclipses definitely show begin and end not only in EA-type (Algol) systems (Samus et al. 2017), but also in EB-type ($\beta$ Lyr) and even EW-type (W UMa) systems (Andronov 2012a, Tkachenko et al. 2016). Thus it is natural either to apply non-polynomial splines to a complete phase light curve, or to make local approximations near the extrema.

Contrary to partial eclipses with relatively smooth variations, during the transits of smaller stars or especially exoplanets, we introduces non-polynomial spline approximations with interval split at (unknown) moments of the inner contacts. This significantly inproves the quality of approximations using the analytical functions.

After numerical tests, we conclude that, for the flat minimum, the best function is WSL ("Wall-Supported Line"), which corresponds to the minimal error estimates either in time, or magnitude. The timing accuracy is better for a "full" data set ($0.000188^d$), which is by a factor of 12 better than the original estimate presented in the discovery paper: Min II HJD 2455506.3590±0.0023 (Devlen, 2015) using the old method of Kwee & van Woerden (1956).

The simple "wall" function $W_1(\varsigma) = (\varsigma^{1.5})_+$ for WSL is adequate for the "short" series with data points at the very bottom of ellipse. but, for "long" series, one may see a distinct change of the sign of the second derivative, thus one has to add one more parameter. We propose to use a "corrected" function $W_2(\varsigma) = (\varsigma^{1.5} \cdot (1 + \lambda \varsigma^2))_+$.

However, for the middle part corresponding to the total eclipse, there may be a variety of functions $V(u)$, and it is possible to choose the best in the program.

These methods are "turned to" symmetrical eclipses. For the pulsating variables, see Andronov et al. (2014).

### 5. Conclusions

The described algorithms are realized in the software MAVKA, which is open for further improvements. It may be effectively used for determination of characteristics of individual minima of eclipsing binary stars with a possibility to choose the method providing the best accuracy for a given data set.